\documentclass[prb,twocolumn,superscriptaddress,showpacs,floatfix]{revtex4}
\usepackage{graphics}
\usepackage{graphicx}
\usepackage{color}

\begin{document}

\title{Intervalley Coupling for Interface-Bound Electrons in Silicon: \\ An Effective Mass Study}
\author{A.L. Saraiva}

\affiliation{Instituto de F\'{\i}sica, Universidade Federal do
Rio de Janeiro, Caixa Postal 68528, 21941-972 Rio de Janeiro,
Brazil}

\author{M.J. Calder\'on}

\affiliation{Instituto de Ciencia de Materiales de Madrid (CSIC), Cantoblanco,
28049 Madrid, Spain}
\author{Rodrigo B. Capaz}

\affiliation{Instituto de F\'{\i}sica, Universidade Federal do
Rio de Janeiro, Caixa Postal 68528, 21941-972 Rio de Janeiro,
Brazil}

\author{Xuedong Hu}

\affiliation{Department of Physics, University at Buffalo, SUNY, Buffalo, NY 14260-1500}

\author{S. Das Sarma}

\affiliation{Condensed Matter Theory Center, Department of Physics,
University of Maryland, College Park, MD 20742-4111}

\author{Belita Koiller}

\affiliation{Instituto de F\'{\i}sica, Universidade Federal do Rio de
Janeiro, Caixa Postal 68528, 21941-972 Rio de Janeiro, Brazil}

\date{\today}

\begin{abstract}

Orbital degeneracy of the electronic conduction band edge in
silicon is a potential roadblock to the storage and
manipulation of quantum information involving the electronic
spin degree of freedom in this host material. This difficulty
may be mitigated near an interface between Si and a barrier
material, where intervalley scattering may couple states in the
conduction ground state, leading to non-degenerate orbital
ground and first excited states. The level splitting is
experimentally found to have a strong sample dependence,
varying by orders of magnitude for different interfaces and
samples. The basic physical mechanisms leading to such coupling
in different systems are addressed here. We expand our
recent study based on an effective mass approach, incorporating 
the full plane wave (PW)  expansions of the Bloch
functions at the conduction band minima. Physical
insights emerge naturally from a simple Si/barrier model. In
particular, we present a clear comparison between ours and
different approximations and formalisms adopted in the
literature, and establish the applicability of these
approximations in different physical scenarios.

\end{abstract}
\maketitle
\pagebreak

\section{Introduction}

Electronic spins in Si are promising candidates for
qubits due to their naturally long coherence times.~\cite{sarma04} An
important challenge for using electronic spins as qubits
in silicon devices is to assure that the low energy physics
is ruled solely by the 2-level spin degree of freedom.~\cite{rahman09, culcer10a,culcer10b,debernardi10,li10,wang10,koenraad11,escott10,rahman11, raith11} In bulk Si crystal, the conduction band lower edge is six-fold
degenerate. Any superposition of the six Bloch states associated with the minima in $k$ along the $\pm x$, $\pm y$ and $\pm z$ crystallographic directions is also an eigenstate of the crystalline hamiltonian, so that the orbital state of a free conduction electron at the band edge is normally not defined. 

Several quantum computer architectures under investigation involve
manipulation of the electronic spin at an interface between Si
and some barrier material, most commonly SiGe alloys~\cite{friesen03,eriksson04} and SiO$_2$~\cite{kane98,kane00} barriers. The confining electric
field (generated by external electrostatic gates and/or
depletion layers) generates a quasi-triangular potential well
at the interface. Assuming the interface to be perpendicular to
the $z$ direction [that is, a (001) interface], the ground
state energy of such triangular potential well depends on the
effective mass in the $z$ direction. The effective masses at
the conduction band minima of Si are anisotropic, with the
longitudinal effective mass more than 4 times larger than the
transversal. This shifts the minima along the $x$ and $y$
directions well above the minima in the $z$ direction, breaking
the six-fold degeneracy into a 2-fold degenerate ground state
and a 4-fold degenerate excited state.~\cite{ando82,kane00PRB}
The splitting is further enhanced if tensile strain is applied
to the Si crystal (\textit{e.g.} in Si quantum wells grown over
relaxed SiGe substrates).~\cite{herring56} The degeneracy in
the two-dimensional $\{k_z,~k_{-z}=-k_z\}$ subspace is lifted
in the presence of a sufficiently singular perturbation
potential, such as a Si/barrier interface.~\cite{saraiva09} Experimental values of the valley splitting in interfaces have been reported in the 0.1 to 1 meV
range.~\cite{ando82,goswami07} A peculiar result
was reported by Takashina {\it et al.},~\cite{takashina06} who
measured a ``giant'' splitting of 23 meV on a Si/SiO$_2$
interface in a SIMOX (separation by implantation of oxygen)
structure.

In the presence of orbital degeneracy, electron
manipulations relying on the Pauli's exclusion principle,
such as Heisenberg exchange coupling~\cite{loss98} and spin blockade~\cite{ono02},
may become unreliable since the qubits Hilbert Space is
now spanned by the valley as well as the spin degrees of freedom.
For an electron spin qubit confined in a Si quantum dot,
reliable knowledge of the interface induced valley splitting
is crucial.~\cite{friesen03}  For example, for a single electron spin
qubit,~\cite{loss98} if the valley splitting is smaller than the reservoir thermal energy, or valley splittings are different across two
quantum dots, exchange gates cannot be performed as was
originally designed since two-valley two-electron singlet
and triplet states (where the two electrons are in different
valleys) are not exchange-split.~\cite{li10}  When
two-electron singlet and triplet states are used to encode a
logical qubit, reliable initialization becomes impossible if
valley splitting in a quantum dot is unknown or known to be
small (compared to reservoir temperature energy scale).~\cite{culcer09}
In short, clear knowledge of a large valley splitting in
a Si quantum dot is imperative to assure the feasibility of
an electron spin qubit.

Early theoretical studies on Si valley splitting in the framework
of the EMA were performed more than 30 years ago.~\cite{ohkawa77,sham79}
The relevance of the periodic part of the Si bulk Bloch states
(leading to the so-called Umklapp processes) to this scattering became
clear in the 70's.~\cite{resta77} However, its inclusion combined with the
EMA formalism leads to a puzzling dependence of calculated physical properties on the particular position of the barrier within the
range of separation between atomic layers, an artifact already
found in previous works~\cite{sham79} which is discussed in
detail and clarified in Ref.~\onlinecite{saraiva09}. These early
theoretical studies do not discuss the
strong sample dependence observed experimentally. 

Valley mixing has been extensively studied in the literature~\cite{ando891,ando892,ando95,menchero99} regarding GaAs/AlAs or GaAs/GaAlAs interfaces and superlattices. The most relevant valley mixing in this context involves the $X$-valleys of the electron conduction band in the AlAs layer, which is the global band minimum in AlAs and a local minimum in GaAs, and the $\Gamma$-valley in the GaAs layer. This mixing involves separated spatial layers. Higher $\Gamma$ and $X$ states in AlAs and GaAs also intervene in the mixing~\cite{ando892}. In Si, mixing occurs between the crystalline momenta $k_z$ and $-k_z$ in the same spatial layer, that is, in the Si slab, and is due to the barrier potential alone. The theory and experimental consequences of GaAs/AlAs valley mixing are thus different and not transferrable to Si.

Performing a fully {\textit ab initio} treatment is not realistic in the study of valley coupling in Si due to limitations both on the length and on the energy scales.
Intervalley splittings of the order of tenths of meV would not
be accurately resolved within the density functional theory (DFT)
approach based on current computational resources. Also, the electronic
states under study spread over several lattice parameters, and
simulation of large supercells with appropriate description of
the band gap (through Hedin's GW scheme,~\cite{hedin65} for
example), again involve numerical computations beyond currently
available capability.

More recent investigations often employ approaches with
atomistic ingredients, whether based completely on the
tight-binding (TB) method or on the hybrid of TB and
EMA.~\cite{grosso96,boykin041,boykin042,friesen06,nestoklon06,lee06,kharche07,boykin08,srinivasan08}
These atomistic methods allow treatment of disorder effects
directly. For example, TB and EMA+TB calculations conclude
that for tilted Si/SiGe quantum wells, alloy disorder and
interfacial step disorder must be included to obtain finite
valley splittings.~\cite{friesen06,kharche07} Such
sample-dependent results are consistent with the observed
variation of this coupling as measured in different interfaces.
On the other hand, the detailed description of disorder comes
at the expense of generality and prevents analytical insights
since these methods require numerical treatment (with exceptions,
some of which are discussed in Sec.~\ref{sec:atomistic}).

We have recently performed~\cite{saraiva09} a study of the valley splitting
problem that leads to the identification of relevant physical
mechanisms underlying the intervalley coupling. Our EMA model
incorporates the atomistic Bloch functions obtained from
\textit{ab initio} calculations in connection with an envelope
function obtained from the single valley EMA equation. That
study focused on the relevance of the barrier height and the
interface width.

In the present work, we discuss the \textit{ab initio} results which are required in implementing our approach. In particular, we provide a detailed roadmap to the connection between effective mass and and the \textit{ ab initio} wavefunction, deriving the formalism and discussing the physics and applicability of our methodology. Further results are presented and discussed, including a more complete analysis of the role of the confining electric field perpendicular to the interface plane. The plane wave (PW) expansion of the periodic part of each Bloch function is explicitly given, and contributions from reciprocal lattice vectors in the PW expansion are analyzed, identifying the most relevant Umklapp processes. We also discuss the present model in the context of existing studies and compare our results with previous ones whenever warranted. The interface is modeled here within EMA by a finite height step potential. Other interface profiles were already considered in our previous study.~\cite{saraiva09}

In section~\ref{sec:theory} we briefly review the model, emphasizing the assumptions involved in the derivation of the formalism. We proceed to numerically calculate the coupling under various electric fields and conduction band offsets in section~\ref{sec:numerical}. These results allow comparison of
the valley coupling in different experimental conditions, and
to discuss possible mechanisms contributing to the giant
splitting reported in Ref.~\onlinecite{takashina06}. Comparison
of our theoretical approach with others in the literature is
given in section~\ref{sec:compare}, where we also evaluate some
of the approximations adopted in previous works and shed light
in some theoretical questions, such as the range of
applicability of various EMA treatments. Finally, our
conclusions are presented in section~\ref{sec:conclusions}.
\section{Theoretical Background}
\label{sec:theory}

We consider a single electron at the bottom of the Si conduction band, near a
(001) Si/barrier interface. Assuming translational symmetry
in the $xy$ plane, we model the barrier material as an
effective potential mimicking the conduction band offset along
the $z$ direction. This model addresses only the position in
energy of the bottom of the conduction band, disregarding the
detailed electronic structure in the transition region between
Si and the barrier material, as we discuss below.

The Hamiltonian for the conduction band electron is assumed to be of
the form

\begin{equation}
H = H_0 + U(z) - e \frac{F}{\epsilon(z)} z,
\label{eq:single-hamilton}
\end{equation}
where $H_0$ is the unperturbed bulk Si Hamiltonian and $U(z)$ is
the barrier potential. An electric field $F/\epsilon(z)$ along
the $z$ direction keeps the electron close to the interface.
The dielectric function $\epsilon(z)$ changes from the bulk Si
value to the bulk barrier material value and, in a more
accurate analysis, could include many-body effects.

A sequence of approximations are involved in obtaining the EMA
equation for Si heterostructures. The first assumption involves
the expansion of the electronic wavefunction in terms of the
periodic part of Bloch states around the conduction band
minima, assumed to be the same for both the well semiconductor
and the barrier material.~\cite{bastard} The envelope function
approach is better justified if the wavevectors at the minima
of the two materials are relatively near relative to the
Brillouin zone dimensions.

We further assume their effective
masses to be the same, although it is possible within EMA to
account for different effective masses in
heterostructures.~\cite{bastard} We avoid specifying the
barrier material and systematically investigate the effects of
the barrier height alone.

Finally, the crystalline structure is
assumed to be preserved across the interface. In a MOSFET
geometry, SiO$_2$ grown over Si is typically amorphous, as
different metastable crystalline phases coexist, with
crystallographic directions that do not necessarily match the
directions of the Si slab.~\cite{helms94} Moreover, the most
energetically favorable crystalline phases of SiO$_2$ have a
single conduction band minimum at the $\Gamma$
point.~\cite{ramos04} On the other hand, Si$_{(1-x)}$Ge$_x$
alloys at low concentrations of Ge ($x<0.3$) present conduction
band minima at the same position in the Brillouin zone as pure
Si crystal,~\cite{rieger93} and the effective mass remains
almost unchanged for the Ge concentration in the alloy up to
$x\sim 0.3$. For samples under experimental
investigation,~\cite{friesen03} Si is epitaxially grown over a
relaxed SiGe substrate, so that the crystallographic directions
match. Although the conduction band states are not exactly the
same for these two materials, they are expected to be very
similar. Thus, for Si/SiGe heterostructures the EMA assumptions
are better justified than for MOSFETs.

However, given the large
conduction band offset between Si and SiO$_2$ a very small
penetration of the envelope function into the barrier material
is expected. Detailed simulation of the electronic structure of
Si/SiO$_2$ interfaces is possible, but does not lead to general
results, since different nanofabrication methods and small
variations of the growth parameters lead to very different
interface morphologies.~\cite{takashina06,takahashi99} We adopt
the EMA approach in the case of Si/SiO$_2$ interface bearing in
mind that this approximation could lead to quantitative
inaccuracies that should be estimated using some other
approach.

Within the above assumptions, one obtains the envelope
function $\Psi(z)$ from the single-valley effective mass
equation~\cite{bastard}
\begin{equation}
\left\lbrace \frac{-\hbar^2}{2 m_z} \frac{\partial^2}{\partial z^2} + U(z) - e\frac{F}{\epsilon(z)} z \right\rbrace \Psi(z) = E \Psi(z),
\label{eq:eff_mass_eq}
\end{equation}
where $m_z$ is the longitudinal effective mass for Si. The
electronic eigenstates of Eq.~(\ref{eq:single-hamilton}) bound
to the interface are obtained from the single-valley
EMA~\cite{kohn} as $\phi_\mu (\mathbf{r}) = \Psi(z) e^{i k_\mu
z} u_\mu (\mathbf{r})$ where $\mathbf{k}_\mu= \pm k_0 \hat z$
are the Bloch wave vectors of the conduction band minima
($k_0\approx0.84\times 2\pi/a_0$). It is convenient to perform
a PW expansion of the periodic functions $u_\pm
(\mathbf{r})$~\cite{koiller04} leading to the complete
wavefunctions

\begin{equation}
\phi_\pm (\mathbf{r}) = \Psi(z) e^{\pm i k_0 z}
\sum_{\mathbf G} c_\pm (\mathbf{G}) e^{i \mathbf{G} \cdot
\mathbf{r}},
\label{eq:expand}
\end{equation}
where $\{\mathbf{G}\}$ are reciprocal lattice vectors. The PW expansion
in equation~(\ref{eq:expand}), originally explored  in
connection to Ref.~\onlinecite{koiller04}, is useful in several 
other contexts.
It was obtained from
{\it ab initio} Density Functional Theory (DFT)
calculations,~\cite{koiller04} performed with the ABINIT
code.~\cite{gonze02} From a DFT perspective, the electronic
correlations for the bulk Bloch states are described in the
Local Density Approximation (LDA).~\cite{hohenberg64,kohn65}
The exchange-correlation potential parameterized by Perdew and
Zunger~\cite{perdew81} from Ceperley-Alder quantum Monte-Carlo
results for the homogeneous electron gas~\cite{ceperley80} was
adopted. The interactions between valence electrons and ions
are described by the {\it ab-initio}, norm-conserving
pseudopotentials of Troullier-Martins,~\cite{troullier91}
generated with the FHI98PP code.~\cite{fuchs99} These
approximations significantly speed up the computation of the
conduction band structure, with a PW expansion of the
wavefunctions including terms up to 16 Ry, that is,
corresponding to 290 plane waves for each $k_\mu$, with
virtually no computational effort.
The calculated equilibrium lattice constant of Si at
$a_0=5.41$ \AA~ and the conduction-band minima at $k_0=0.844
(2\pi/{\rm a_0})$ are in close agreement with experimental
results.~\cite{madelung} These values are used in the
calculations presented below.

We find that over 90$\%$ of the spectral weight of the PW expansion
in equation~(\ref{eq:expand})
comes from the five points in the BCC reciprocal lattice which
are nearest to each conduction band minimum $k_\mu$.
The valley coupling
\begin{equation}
V_{VO} = \langle\phi_+|H|\phi_-\rangle
\label{eq:vvo}
\end{equation}
is the key quantity leading to the valley splitting $\Delta = 2
|V_{VO}|$.~\cite{saraiva09}
Therefore,
a preliminary estimate of the $\pm z$ intervalley coupling could involve 9 PWs (the $\Gamma$
point is a common nearest neighbor for both $k_\mu$ at the
minima). The coefficients such that $|c_\mu (\mathbf{G})|^2>10^{-4}$, accounting for 99.5 \% of the spectral weight, are explicitly given for $\mu = z$ in Table \ref{tab:coefs}; coefficients for all band minima may be obtained from those by symmetry (see Table caption).
\begin{table}[ht!]
\begin{center}
\begin{tabular}{|c||c|c||c|} \hline
${\mathbf G}$ & Re$[c_{+z}({\mathbf G})]$ & Im$[c_{+z}({\mathbf G})]$ & $|c_{+z}({\mathbf G})|^2$ \\ \hline \hline
( 1 -1 -1) & -0.3131 & -0.3131 & 0.1961\\ \hline
(-1  1 -1) & -0.3131 & -0.3131 & 0.1961\\ \hline
( 1  1 -1) & -0.3131 & 	0.3131 & 0.1960\\ \hline
(-1 -1 -1) & -0.3131 & 	0.3131 & 0.1960\\ \hline
( 0  0  0) &  0.3428 & -0.0000 & 0.1175\\ \hline
( 2  0 -2) & -0.0986 & 	0.0000 & 0.0097\\ \hline
( 0  2 -2) & -0.0986 & 	0.0000 & 0.0097\\ \hline
(-2  0 -2) & -0.0986 & 	0.0000 & 0.0097\\ \hline
( 0 -2 -2) & -0.0986 & 	0.0000 & 0.0097\\ \hline
( 1 -1  1) &  0.0695 & -0.0695 & 0.0097\\ \hline
(-1  1  1) &  0.0695 & -0.0695 & 0.0097\\ \hline
( 1  1  1) &  0.0695 & 	0.0695 & 0.0097\\ \hline
(-1 -1  1) &  0.0695 &  0.0695 & 0.0097\\ \hline
(-2  2 -2) & -0.0000 & -0.0451 & 0.0020\\ \hline
( 2 -2 -2) & -0.0000 & -0.0451 & 0.0020\\ \hline
(-2 -2 -2) &  0.0000 & 	0.0451 & 0.0020\\ \hline
( 2  2 -2) &  0.0000 & 	0.0451 & 0.0020\\ \hline
( 0  2  2) &  0.0387 & -0.0000 & 0.0015\\ \hline
( 2  0  2) &  0.0387 & -0.0000 & 0.0015\\ \hline
( 0 -2  2) &  0.0387 & -0.0000 & 0.0015\\ \hline
(-2  0  2) &  0.0387 & -0.0000 & 0.0015\\ \hline
( 0  0 -4) &  0.0186 & -0.0000 & 0.0003\\ \hline
(-1  1  3) &  0.0114 & 	0.0114 & 0.0003\\ \hline
( 1 -1  3) &  0.0114 & 	0.0114 & 0.0003\\ \hline
(-1 -1  3) &  0.0114 & -0.0114 & 0.0003\\ \hline
( 1  1  3) &  0.0114 & -0.0114 & 0.0003\\ \hline
( 0  0  4) &  0.0121 & -0.0000 & 0.0001\\ \hline
( 3 -3 -1) & -0.0075 & -0.0075 & 0.0001\\ \hline
(-3  3 -1) & -0.0075 & -0.0075 & 0.0001\\ \hline
( 3  3 -1) & -0.0075 &  0.0075 & 0.0001\\ \hline
(-3 -3 -1) & -0.0075 & 	0.0075 & 0.0001\\ \hline
\end{tabular}
\caption{\label{tab:coefs} Plane wave expansion coefficients $c_\mu (\mathbf{G})$ for $\mu = +z$. The integers in the first column give $G_1,~G_2,~G_3$, respectively the  $(x, y, z)$ $\bf G$ cartesian coordinates in units of $(2\pi/{\rm a})$. The real and imaginary parts of $c_{+z} (\mathbf{G})$ are displayed in columns 2 and 3 respectively. Column 4 shows $|c_{+z} (\mathbf{G})|^2$, and only coefficients  $|c_{+z}|>10^{-2}$ are shown. The coefficients for all other minima may be obtained using the symmetry relations $c_\mu (\mathbf{G}) = c^*_{-\mu} (\mathbf{-G})$; $c_x(G_1,G_2,G_3) = c_z(G_3,G_2,G_1)$, and  $c_y(G_1,G_2,G_3) = c_z(G_1,G_3,G_2)$.}
\end{center}
\end{table}

From the above expansion, eq. (\ref{eq:vvo}) reads
\begin{equation}
V_{VO}=\sum_{\mathbf{G}, \mathbf{G^\prime}} c^*_+(\mathbf{G})
c_-(\mathbf{G^\prime}) \delta(G_x-G^\prime_x)\delta
(G_y-G^\prime_y) I(G_z,G'_z),
\label{eq:sum}
\end{equation}
where the last term stands for the integral
\begin{equation}
I (G_z,G'_z)= \int_{-\infty}^{+\infty} |\Psi(z)|^2 e^{i Q z} \left[U(z)- \frac{F}{\epsilon(z)} z\right] \rm{d}z\\
\label{eq:integral}
\end{equation}
with $Q= G_z - G^\prime_z -2 k_0$.

We take for $U(z)$ the step potential, which is the most
favorable model for the conduction band profile between Si
and the barrier in terms of maximizing the valley coupling.~\cite{saraiva09} It
is written as~\cite{bastard}
\begin{equation}
U(z) = U_{{\rm step}}(z) = U_0 \Theta (z-z_I),
\label{eq:step}
\end{equation}
where $z_I$ is the position of the interface and $U_0$ is the
conduction band offset. The step potential aims at modeling a
perfectly sharp interface, a concept that involves changing the
species of the atomic constituents abruptly, i.e. across one
monolayer (1ML). However, the envelope function equation allows
a continuous choice of the interface position $z_I$, even
within 1 ML width, as discussed in Ref.~\onlinecite{saraiva09}.
Although $z_I$ is ill-defined within a ML length scale, it is
convenient to keep this simple interface model for it allows
decomposing the integral in equation~(\ref{eq:integral}) into
terms that are easily identifiable, some of which are familiar
from previous theoretical treatments. Other models of
the interface have been studied in Ref.~\onlinecite{saraiva09}.

In equation~(\ref{eq:integral}) we integrate by parts the term
proportional to $U(z)=U_{\rm step}(z)$ given in
(\ref{eq:step}),
%
\begin{eqnarray}
I (G_z,G'_z) &=& \frac{i U_0}{Q} \int_{-\infty}^{\infty} \delta(z-z_I) \left|\Psi(z)\right|^2 e^{i Q z} {\rm d} z \nonumber\\
             &+& \frac{i U_0}{Q} \int_{-\infty}^{\infty}  \Theta (z-z_I)\frac{\rm{d}|\Psi(z)|^2}{\rm{d}z} e^{i Q z} {\rm d} z\nonumber \\
             &-& \int_{-\infty}^{\infty} \left|\Psi(z)\right|^2 e^{i Q z} \frac{F}{\epsilon (z)} z {\rm d}z\nonumber\\
             &=& \frac{i}{Q} U_0 \left|\Psi(z_I)\right|^2 e^{iQz_I}  \label{eq:integraldelta} \\
             &+& \int_{z_I}^{\infty} \frac{i}{Q} U_0 \frac{\rm{d}|\Psi(z)|^2}{\rm{d}z} e^{i Q z} {\rm d} z \label{eq:integrale} \\
             &-& \int_{-\infty}^{\infty} \left|\Psi(z)\right|^2 e^{i Q z} \frac{F}{\epsilon (z)} z {\rm d}z~, \label{eq:integralf}
\end{eqnarray}
%
where we are interested in an interface bound state, so that
the electronic density $\left|\Psi(z)\right|^2$ vanishes at
$z\rightarrow\pm\infty$. Three terms are left, labeled
(\ref{eq:integraldelta}), (\ref{eq:integrale})
and~(\ref{eq:integralf}) above. Term~(\ref{eq:integraldelta})
evidences the role of the electronic density at the interface
$\left|\Psi(z_I)\right|^2$ through a $\delta$ function at
$z=z_I$; term~(\ref{eq:integrale}) gives the contribution of
the evanescent tail of the electronic envelope function into
the barrier material $z>z_I$; term~(\ref{eq:integralf})
represents an  intervalley scattering induced directly by the
electric field, which we find to be vanishingly small.
Summation of these contributions over the reciprocal lattice
vectors [see equation~(\ref{eq:sum})] leads to three
contributions to the intervalley coupling
\begin{equation}
V_{VO}=V_\delta+V_E+V_F,
\label{eq:three_terms}
\end{equation}
that is, the delta-function contribution $V_\delta$, the
evanescent term contribution $V_E$, and the electric field
contribution $V_F$.

Besides the mismatch between the conduction band minima of the
two materials, we also consider the change in dielectric
screening constant from the semiconductor to the insulator. The
dielectric function is taken as $\epsilon(z< z_I)=\epsilon_{\rm
Si}$ and  $\epsilon(z> z_I)=\epsilon_{\rm barrier}$.
%
Note that $\epsilon(z)$ also introduces a \textit{kink} in the
electrostatic potential in the scale of the monolayer
separation, which could in principle contribute to the
intervalley coupling.

\section{Numerical Solution and Results}
\label{sec:numerical}

\subsection{Envelope Function Variational Approaches: Finite Differences and Trial Function}
\label{sub:envelope}

The envelope function $\Psi(z)$ is obtained from equation~(\ref{eq:eff_mass_eq}), which has no analytic solution. The approach adopted to solve this equation must be carefully chosen, since the intervalley coupling depends explicitly on the envelope function details, as shown in equations~(\ref{eq:integraldelta}), (\ref{eq:integrale}) and (\ref{eq:integralf}). The first term~(\ref{eq:integraldelta}), which gives rise to $V_\delta$ in equation~(\ref{eq:three_terms}), is proportional to the electronic probability density at the interface$\left|\Psi(z_I)\right|^2$, while the second term~(\ref{eq:integrale}) depends on the envelope tail inside the barrier, defining the contribution $V_E$ of the evanescent tail. The term~(\ref{eq:integralf}) is always found to be negligibly small compared to the others, and is not included in the results presented here.

In order to get an analytic approximation for the envelope function, we present initially results obtained using the variational approach; we tried several functional forms for the variational envelope function, all satisfying the following boundary conditions: in the far semiconductor region ($z \rightarrow - \infty$), under a constant electric field, the envelope function is approximated by a gaussian decay; towards the barrier material the short range decay ($z \mathrel{\rlap{\lower4pt\hbox{\hskip1pt$\sim$}}\raise1pt\hbox{$>$}} z_I$) is nearly exponential.
The trial function that typically gave the lowest energy expectation value was
\begin{equation}
\Psi(z)=\left\{ \begin{array}{ll}
                     \Psi_A={\cal A} N_A (z_0-z) e^{-\alpha(z_0-z)^2} &, z< z_I\\
                     \Psi_B= {\cal B} N_B e^{-\beta z}                &, z>z_I\\
                    \end{array}
            \right.
\label{eq:trial-parts}
\end{equation}
with $N_i=\left(\int dz\,\, \Psi_i^2 \right)^{-1/2}$ (i=A,B).

If we take the barrier to be infinite, the optimized variational parameters become $z_0=z_I$ and ${\cal B}=0$. But in general the barriers are finite, some penetration is expected so that $z_0>z_I$ and ${\cal B}>0$. From the continuity conditions $\Psi_A(z_I)=\Psi_B(z_I)$ and $\Psi'_A(z_I)=\Psi'_B(z_I)$, and the normalization ${\cal A}^2+{\cal B}^2=1$, we obtain expressions for $\cal A$, $\cal B$, and $z_0$ in terms of the variational parameters $\alpha$ and $\beta$; e.g. For $z_I=0$, we have

\begin{equation}
z_0=\frac{-\beta+\sqrt{8 \alpha+\beta^2}}{4 \alpha}
\label{eq:z0}
\end{equation}

\begin{equation}
{\cal A}=\left(\sqrt{\frac{1}{\beta}} \sqrt{\beta+\frac{8 z_0^2 \,\alpha^{3/2}}{4 z_0 \sqrt{\alpha}+e^{2 z_0^2 \alpha} \sqrt{2 \pi}\, {\rm erfc}(\sqrt{2 \alpha} z_0)}}\right)^{-1}
\label{eq:A}
\end{equation}

\begin{figure}[ht!]
\resizebox{90mm}{!}{\includegraphics{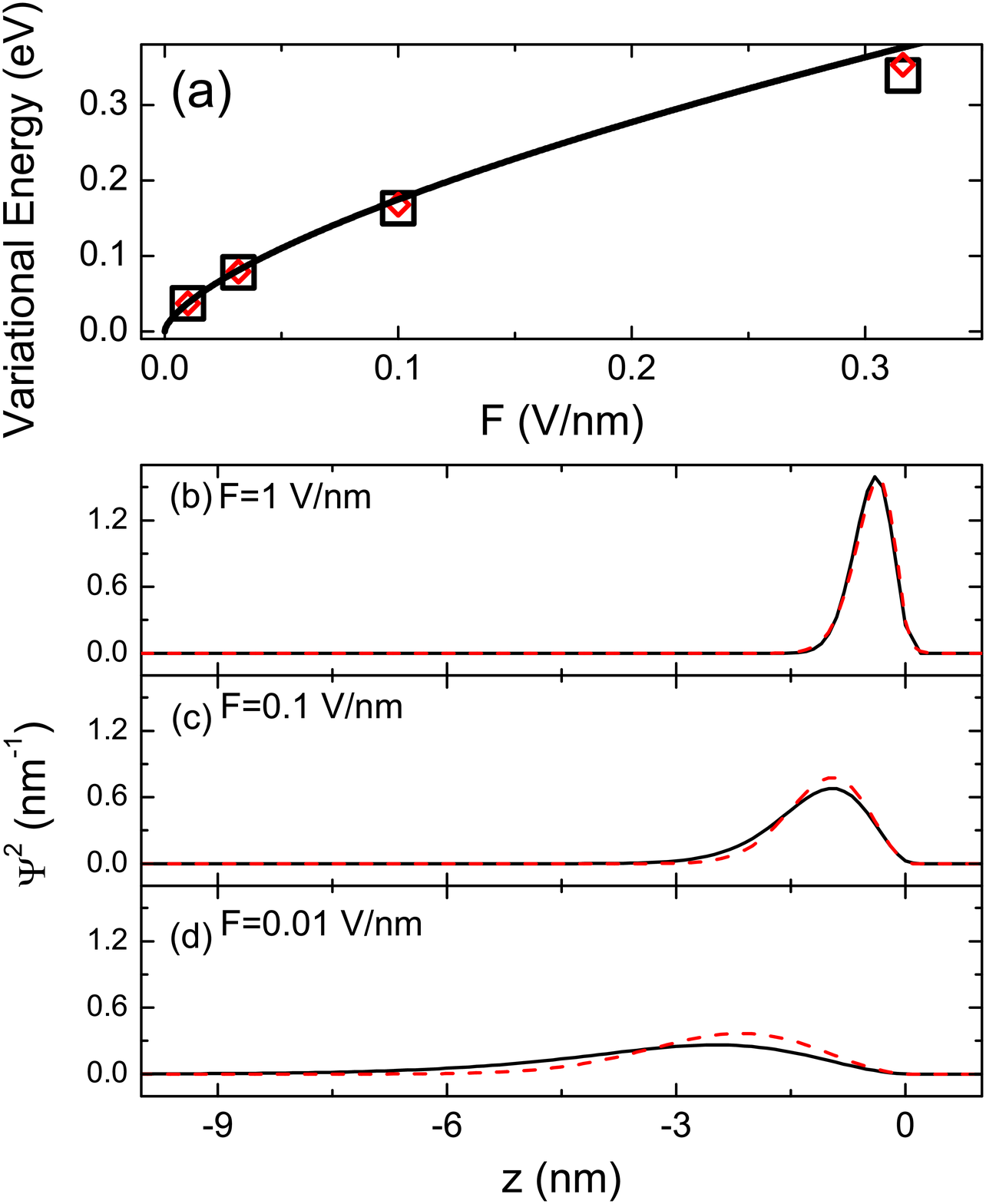}}
\caption{(Color online) Energies and envelope functions for a conduction band offset $U_0=$ 3 eV. (a) Comparison between the expectation value for the ground state energies obtained variationally by solving the Finite Differences equation through the steepest descent method (squares) and using the trial wavefunction  in equation~(\ref{eq:trial-parts})  (diamonds). Exact results for $U_0\to\infty$ from Ref.~\onlinecite{stern72}, are also shown (solid line). (b)-(d) Envelope functions squared (probability densities) for the indicated electric fields.
The solid line is obtained solving the Finite Differences equation variationally (through the steepest descent method). The dashed line is obtained from the trial-function in equation~(\ref{eq:trial-parts}).
In (b), $F=1.00$~V/nm and the two approaches are in good agreement, while
at lower electric fields, (c) and (d), the
functional form in equation~(\ref{eq:trial-parts}) leads to somewhat more localized states near the interface. }
\label{fig:envelopes}
\end{figure}

\begin{equation}
{\cal B}= \sqrt{\frac{8 z_0^2 \, \alpha^{3/2}}{4 z_0 \sqrt{\alpha}(2 z_0 \alpha+\beta)+e^{2 z_0^2 \alpha} \sqrt{2 \pi}\, \beta \, {\rm erfc}(\sqrt{2 \alpha} z_0)}}
\label{eq:B}
\end{equation}
where erfc is the complementary error function,
\begin{equation}
{\rm erfc}(x)=1-{\rm erf}(x)=1-\frac{2}{\sqrt{\pi}} \int^x_0 e^{-t^2} dt.
\end{equation}
Analytic solutions for the case of $z_I\ne 0$ may also be obtained. However, for the step function potential and at relatively large length scales, the interface position is irrelevant: it separates two semi-infinite regions, one filled with Si and the other with the barrier material. At atomic distance length scales, the coupling dependence on $z_I$ shows oscillatory behavior with a period of 1 ML, a peculiarity  of the EMA combined with the underlying Bloch states, as discussed in detail in Ref.~\onlinecite{saraiva09}. Data presented there (Figure 5 of Ref.~\onlinecite{saraiva09}) indicate that, for the step potential, these oscillations cause an uncertainty of about $20\%$ in $|V_{VO}|$. In what follows, for simplicity and definiteness, we arbitrarily fix the value of $z_I = 0$ bearing in mind that our results for the coupling are not accurate within up to $\sim 20\%$.

In order to provide a more robust estimate of the wavefunction, independent of our choice of its functional form, we also solve equation~(\ref{eq:eff_mass_eq}) through a finite
differences method.
This approach allows us to obtain the wave function
variationally without inputing a guess for the functional form of the
envelope. The envelope function is discretized and the
derivatives are approximated by the slope of the interpolated
straight line between two discrete points. Each discrete point
is taken as a separate variational parameter, under the constraint
that the wave function is normalized. The minimum energy in this
configuration space is obtained through the Steepest Descent method.
This strategy should in principle permit us to
discuss a wide range of electric fields.

The trial function and the finite differences expectation value of the energy are given in Fig.~\ref{fig:envelopes}(a), according to which the lowest value down to $F \sim 0.01$ V/nm is obtained within  finite differences/steepest descent method.  The exact result for infinite barrier, given in Ref.~\onlinecite{stern72}, is also included, and a good agreement with our numerical results is obtained. As the field increases and  pushes the wavefunction towards the barrier, an evanescent tail into the barrier region is formed for $U_0=3$ eV, lowering the electronic energy with respect to the impenetrable barrier $U_0 \to \infty$ case, as expected [upper right data in Fig.~\ref{fig:barrier}(a)]. The good agreement for the expectation value of the energy among the three methods, for example at F=0.1 V/nm in Fig 1(a), does not imply that the wavefunctions are in agreement - as illustrated in Fig~\ref{fig:envelopes}(c) and as is
obvious for the infinite barrier case, where the wavefunction for $z\ge 0$ is exactly zero. It is known that the energy
alone is not a valid criterion to guarantee the wavefunction validity, in particular here for $z \ge 0$ - from which the
coupling is calculated. Imposing a predetermined form to the envelope most certainly will give spurious results for $V_\delta$ and $V_E$. This points the numerical finite differences as the most adequate approach, with the additional capability of describing different interface profiles.~\cite{saraiva09}

The finite differences approach is expected to be more accurate
than trial-function based methods. However, in
practice, numerical constraints limit its implementation
and reliability. Below $F \approx 0.01$ V/nm (not shown), we
find that the lowest energy actually corresponds to Eq.~(\ref{eq:trial-parts}), a consequence of our numerical accuracy limitations. Comparison
between the envelope functions squared (probability
densities) for a range of electric fields and for a conduction
band offset $U_0 = 3$ eV is given in Fig.~\ref{fig:envelopes}(b)-(d). It is
clear from the figures that the envelope defined in Eq.~(\ref{eq:trial-parts})) is in very good agreement with the one obtained from Finite Differences for relatively high electric fields [this is illustrated in Fig.~\ref{fig:envelopes}(b)], but the difference at lower
electric fields is noticeable [see Fig.~\ref{fig:envelopes}(c), (d)].
Results presented below correspond to the Finite Differences
solution to equation (2) in the $F \gtrsim 0.01$ V/nm regime.

\subsection{Contribution from $V_F$, $V_\delta$ and $V_E$}

We now proceed to calculate the three terms in equation~(\ref{eq:three_terms}).
As mentioned above, we find the electric field contribution $V_F$ to be orders of magnitude smaller than the other two terms for any value of
$\epsilon_{\rm{barrier}}$. This is an indication that the
\textit{kink} in the electrostatic potential introduced by the
change in dielectric constants is not singular enough to produce sizable coupling between the $\pm z$ valleys.

The other two terms are also left unchanged if the same
conduction band offset $U_0$ is imposed and different values of
$\epsilon_{\rm {barrier}}$ are adopted, meaning that the
electronic wavefunction in all cases does not penetrate the
barrier material deep enough to be affected by the electric
field inside it. Therefore, we disregard the
electric field in the barrier. This is
beneficial to our model, since we can characterize the barrier
material by the conduction band offset alone, and not involve
other properties specific to the barrier material.

A richer behavior is obtained from $V_{\delta}$,
the contribution from the $\delta$-function given in
Eq.~(\ref{eq:integraldelta}). Since this term is
proportional to the product $U_0 \left|\Psi(z_I)\right|^2$,
there is a trade-off between the barrier height and the
envelope function penetration into the barrier material. In
Fig.~\ref{fig:barrier} we note that $|V_{\delta}|$ increases
with $U_0$, meaning that, for $F=10^{2}$ V/nm,  the increase in $U_0$
prevails over the reduction in $\left|\Psi(z_I)\right|^2$.

Finally, the term $V_E$ arising from the evanescent tail
contribution $\Psi(z>z_I)$ also presents a non-trivial
trade-off. While some penetration of the envelope function
into the barrier material is needed for this term to be
non-vanishing, the integrand in equation~(\ref{eq:integrale})
is highly oscillatory so that if the envelope function
penetrates more than a few monolayers, $V_E$ integrates to
zero. Since SiGe barriers present fairly large electronic
penetration, the calculated $V_E$ for these materials is much
smaller than in the case of SiO$_2$ barriers. Besides, in the
SiGe barrier the $V_E$ contribution only gives a change in the
complex phase of $V_{VO}$, thus leading to $|V_{VO}|\approx |V_{\delta}|$. Still, since this small $V_E$ contribution to
the intervalley coupling in Si/SiGe heterostructures changes
its complex phase, it leads to a different ground state
combination of the $\pm z$ valleys.

\subsection{Umklapp Processes}

Since we take the plane wave expansion
of both $\pm z$ conduction band minima, Umklapp processes are
fully included here. An estimate of their relevance is given by
the total summation over them, which gives $|V_{\delta}|_{\rm Umklapp}= 0.02907\, U_0
\left|\Psi(z_I)\right|^2 a_0$, while the contribution from ${\bf G}={\bf G'} = 0$ is $|V_{\delta}|_\Gamma = 0.01108\, U_0 \left|\Psi(z_I)\right|^2 a_0$.

We also look at each plane wave contribution separately. 
\emph{The most prominent terms in the PW expansion of the Bloch functions
at the conduction band minima are not the} ${\mathbf G}=0$ \emph{term, but the
first nearest neighbors in the BCC reciprocal lattice}, as can be seen in Table~\ref{tab:coefs}.

Table~\ref{tab:contribimag} shows the numerical prefactors, multiplying $U_0 \left|\Psi(z_I)\right|^2 a_0$, of the contributions to the imaginary part, confirming that the most relevant contributions come from the $\Gamma$ point and its first eight neighbors in the BCC reciprocal lattice of Si since we get over 70\% of the total value coming from these points. Notice the Umklapp terms are the most relevant.

It is also interesting to note that for the imaginary part the most relevant contributions are obtained for $\mathbf{G}=\mathbf{G^\prime}$. This is equivalent to taking the zeroth order plane wave expansion of the product $u^*_+(\mathbf{r}) u_-(\mathbf{r})$, a commonly adopted approximation in the EMA theory of shallow donors.~\cite{debernardi06} But a complete analysis of the Umklapp processes reveals that this is particular to the delta contribution, and that there is no general justification for this approximation for the interface induced valley coupling.  

\begin{table}[htb]
\begin{center}
\begin{tabular}{|c|c||c|} \hline
$\mathbf{G} (2\pi/a)$  &  $\mathbf{G}^\prime (2\pi/a)$ & Contribution $\left(U_0\left|\Psi(z_I)\right|^2 a_0\right)$\\ \hline \hline
     $(0,0,0)$         &           $(0,0,0)$           &      -0.01108            \\ \hline
     $(1,1,1)$         &           $(1,1,1)$           &      0.00410            \\ \hline
     $(-1,1,1)$        &           $(-1,1,1)$          &      0.00410            \\ \hline
     $(1,-1,1)$        &           $(1,-1,1)$          &      0.00410            \\ \hline
     $(1,1,-1)$        &           $(1,1,-1)$          &      0.00410            \\ \hline
     $(-1,-1,1)$       &           $(-1,-1,1)$         &      0.00410            \\ \hline
     $(-1,1,-1)$       &           $(-1,1,-1)$         &      0.00410            \\ \hline
     $(1,-1,-1)$       &           $(1,-1,-1)$         &      0.00410            \\ \hline
     $(-1,-1,-1)$      &           $(-1,-1,-1)$        &      0.00410            \\ \hline
\end{tabular} \caption{\label{tab:contribimag}Most relevant contributions to the imaginary part of $V_{\delta}$. The sum of these terms is $0.02175\, U_0 \left|\Psi(z_I)\right|^2 a_0$, underestimating this term by only 25\%. The real part of this contribution vanishes due to our choice of coordinate system.}
\end{center}
\end{table}

\subsection{Complete Valley Coupling }

\begin{figure}
\resizebox{80mm}{!}{\includegraphics{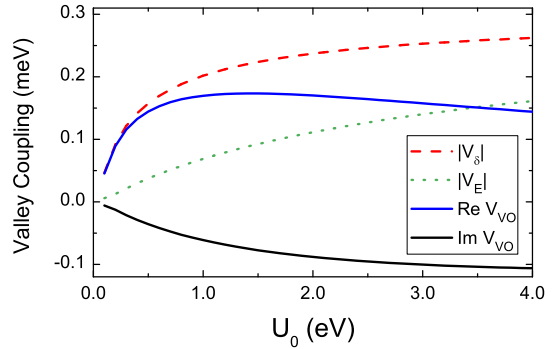}}
\caption{(Color online) Intervalley coupling as a function of the conduction band offset $U_0$.
All data correspond to an external field of 10$^{-2}$ V/nm.
The absolute value of the delta function contribution $|V_\delta|$ and the evanescent tail contribution $|V_E|$ are
depicted separately, as well as the real and imaginary parts of the total coupling $V_{VO}=V_\delta + V_E$.
The offset for Si/SiGe interfaces is $0.1 - 0.2$ eV and for Si/SiO$_2$ interfaces is $3$ eV.}
\label{fig:barrier}
\end{figure}

The absolute values, $|V_\delta|$ and $|V_E|$, of the two terms contributing
to $V_{VO}$ and $|V_{VO}|=|V_\delta+V_E|$ are shown as a function of the barrier
height in Fig.~\ref{fig:barrier}. It is clear here that in general the two terms
have different behaviors, and that they both give important
contributions to the total coupling. However, for estimating the
valley splitting, $|V_{VO}|=|V_{\delta}|$ is a reasonable approximation
for small barriers, such as those in Si/SiGe heterostructures.
Noting that $|V_{VO}|<|V_\delta|+|V_E|$, we infer that
$V_{VO}$ is in general a complex number, not a purely
imaginary quantity~\cite{saraiva09}.
Since $V_\delta$ is purely imaginary, this triangle inequality is only
true if $V_E$ has a non-vanishing real part. For instance, at
$U_0=3$eV, we have an intervalley coupling of $V_{VO}= (-0.100+
i 0.158)$meV. Also, $V_\delta$ and $V_E$ increase monotonically with $U_0$,
while $V_{VO}$ decreases at large offsets. This indicates that the relative
phase between $V_\delta$ and $V_E$ changes with $U_0$.

In principle, SiGe and SiO$_2$ barriers could lead to similar
intervalley couplings. Of course these two materials present
very different interface morphologies and are grown with
different techniques, which should lead to differences between
the intervalley couplings measured in each design. There has been reports of significantly different valley splitting even for the two interfaces of the same quantum well.~\cite{takashina04}

\begin{figure}[ht!]
\resizebox{80mm}{!}{\includegraphics{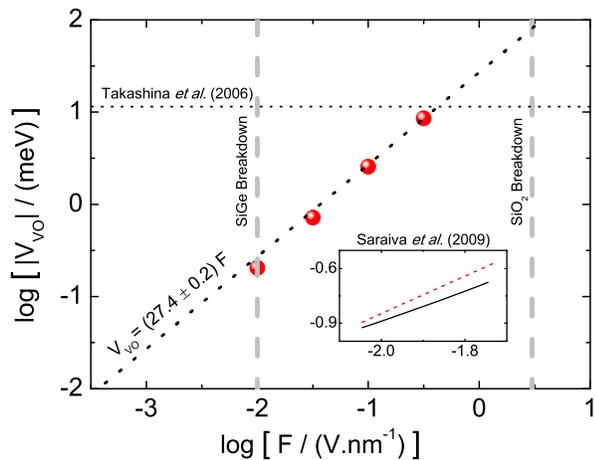}}
\caption{(Color online) Data points give the calculated
intervalley coupling as a function of applied electric field, both
in a log scale, for barrier height $U_0 = 3$ eV.
A linear fit for the data points with $V_{VO}$ given in meV and $F$ in V/nm, is included (dotted line through the points), leading to an estimated $\lambda\sim 0.27$\AA. The vertical  dashed
lines correspond to the experimental values of the
breakdown field for SiGe and SiO$_2$. The dotted horizontal
line represents the measured valley coupling (or half the splitting)
reported in Ref.\onlinecite{takashina06}. The solid straight line is
a possible prediction from the model in Ref.~\onlinecite{sham79},
calculated from equation~(\ref{eq:SeN}), taking $\alpha = 0.43$
\AA\ and $\langle\frac{\partial u}{\partial z}\rangle=F$.
The inset shows the same results presented in Ref.~\onlinecite{saraiva09},
covering a much narrower range of field values, for $U_0 = 3$ eV
(solid line) and 125 meV (dotted line)}.
\label{fig:field}
\end{figure}

The valley coupling calculated for a range of external
electric fields, covering several orders of magnitude are
summarized in Fig.~\ref{fig:field}. This range expands by two orders of magnitude the previously reported range in Ref.~\onlinecite{saraiva09} (see inset). We take the upper limit of $F$ just below  the SiO$_2$ breakdown
field $F \approx 3$ V/nm. This bound is indicated in the figure, as well
as  the SiGe breakdown field $\approx 10^{-2}$ V/nm. The lower value of $F$ is
constrained by our numerical accuracy, as discussed in Sec.~\ref{sub:envelope}.
For large enough values of F, the valley splitting can be of the order of 10 meV, which is compatible (within error) with the giant splitting observed by Takashina \textit{et al.}~\cite{takashina06}
This means that the results obtained in a SIMOX interface could be related to nearly perfectly sharp
interfaces combined with relatively high gate fields, which
might be consistent with the experimental conditions
in terms of these two parameters. Other effects may be present that can further enhance this coupling and account for or contribute to very large splittings, as the recently proposed mechanism involving interface states.~\cite{saraiva10}

Previous studies~\cite{sham79,friesen07} also report a linear
dependence $|V_{VO}| = \lambda F$, where $\lambda$ is a
model-dependent length. This behavior in different models is qualitatively and quantitatively addressed in Sec.~\ref{sec:compare}.

\section{Discussions and Connections with previous model calculations}
\label{sec:compare}

The problem of interface-induced intervalley coupling has been
treated extensively in the literature, particularly within EMA.
Since the complete problem is not exactly solvable, several
approximations regarding the barrier potential and the nature
of the electronic states have been adopted in different
contexts. We discuss in this Section connections between our study and
some previous contributions.

\subsection{Electric Field Dependence of the Valley Splitting}
\label{sec:EMA}

The applied field $F$ is one of the key parameters affecting $V_{VO}$, and probably the most controllable.
Early work by Sham and Nakayama~\cite{sham79} established a linear dependence of the valley splittting on $F$. They address the Si valley splitting
problem from the perspective of the electrons in a space-charge
layer in a MOSFET. The effect of the Si-SiO$_2$
interface on electron dynamics is studied by considering incoming (towards
the interface) and outgoing (away from the interface) Bloch
waves in the Si region and evanescent waves decaying from the
barrier inside the semiconductor. The potential barrier is modeled to be infinite, which
disregards the evanescent tail of the electronic wavefunction
into the barrier. Also, their study is devoted to the case of
a planar density of electrons at the interface (of the order of
$10^{12}$ cm$^{-2}$ electrons), including many body
corrections. In contrast, we develop here a theory for a \emph{single
electron in a bound state}, relevant in the context of quantum computing. Therefore, any comparison with Ref.~\onlinecite{sham79}
should be taken with caution.

The valley splitting obtained by Sham and Nakayama,~\cite{sham79}
\begin{equation}
\Delta E = 2 |V_{VO}| = \left |\alpha \langle \frac{\partial u}{\partial z}\rangle\right|,
\label{eq:SeN}
\end{equation}
is proportional to the mean value of the derivative of the
self-consistent potential. The parameter $\alpha$, a
characteristic length related to the intervalley scattering
matrix, is given as a function of the interface position with
respect to a crystal (001) plane, (called $z_0$ there), showing
sharp variations as $z_0$ runs over a 1 ML range. But the concept
of an interface as an abrupt change in conduction band energy
is not well defined in this scale.
This ambiguity comes from the continuum EMA
approach combined with the atomistic band structure input. These same
ingredients affect our results, with the calculated valley
coupling oscillating as a function of $z_I$.~\cite{saraiva09}
We estimate that the mean value of the derivative
of the self-consistent potential should be of the order of the field, $\langle\frac{\partial
u}{\partial z}\rangle \cong F$, and take $|\alpha|=0.43$ \AA\ as estimated
in Ref.~\onlinecite{sham79}. Therefore $\lambda (S\&N)=\alpha/2 = 0.21$ \AA.

More recent EMA-based models~\cite{friesen06,friesen07,goswami07,chutia08,friesen10}
rely on an effective coupling potential responsible for the
intervalley coupling, where the perturbation potential is
taken as a $\delta$-function with strength obtained from either
TB atomistic calculations or experimental data.
In our formalism, it means taking in equation~(\ref{eq:three_terms}) only the term $V_\delta$. As discussed
in the previous section and shown in Fig.~\ref{fig:barrier},
this approximation is better for the relatively low SiGe barriers, but for barrier
materials with higher conduction band offsets, sizeable contributions (actually reducing $|V_{VO}|$) come from the $V_E$ term.
The valley couplings in Refs.~\onlinecite{friesen07} and \onlinecite{chutia08} are also found to be linear with electric field, and their estimated values for $\lambda$ are 0.72 \AA~and 1.36 \AA~respectively.

In our model calculations, the direct $\{k_z,k_{-z}\}$ coupling ($V_F$) mediated  by the field term in $H$ is found to be negligible. The effectiveness of $F$ comes from the electronic charge being pulled towards the barrier material, increasing the wavefunction amplitude at and beyond the interface position.
These shifts contribute to the terms $V_\delta$ and $V_E$ in (\ref{eq:three_terms}).
From the linear fit of our data in Fig.~\ref{fig:field}, we get $\lambda$(present work)=27.4 mV/(V/nm) = 0.27\AA, thus comparable to $\lambda (S\&N)$.\footnote{Fits to the results in Ref.\onlinecite{saraiva09} (inset of Fig.~\ref{fig:field}) give $\lambda = 0.14$ \AA\ or 0.12 \AA\  for $U_0$ = 150 meV or 3 eV. For $U_0=3$ eV, the value $\lambda = 2.7$ \AA, quoted here, is more reliable since it fits a wider range of fields, with data points obtained from a more accurate finite differences numerical solution.} Results in Ref.~\onlinecite{friesen07} (alternatively, Ref.~\onlinecite{chutia08}) are  larger than $S\&N$ by a factor of 3 (6), and are 2 (5) times larger than ours.  Of course, there are many differences between the models which may account for the different results, but it is interesting to note from Fig.~\ref{fig:barrier} that the delta-function term alone, $|V_\delta|$,  always overestimates the full coupling $|V_{VO}| = |V_\delta + V_E|$. Therefore, considering only the delta function potential in the coupling term may be related to larger values predicted for $\lambda$ in Refs.~\onlinecite{friesen07} and \onlinecite{chutia08}.

The TB approach is also used
to obtain more accurate boundary conditions at the interface
for the EMA wavefunction.~\cite{chutia08} Furthermore,
the ambiguity introduced by the interface position discussed before
was also recognized in this study, and explored in Ref.~\onlinecite{friesen07}
to treat the interface position within the uncertainty range as a fitting parameter to match the TB
and effective mass wavefunctions in a finite quantum well.

\subsection{Atomistic Approaches}
\label{sec:atomistic}

More realistic quantitative estimates may be obtained from
atomistic many band TB description of Si and the
barrier
material,~\cite{grosso96,boykin041,boykin042,friesen06,nestoklon06,lee06,kharche07,boykin08,srinivasan08}
which may also account for interface disorder such as alloying
effects and surface roughness. Such approaches are not always adequate or meant
to give a general picture of the physical mechanisms
behind the intervalley coupling, since the focus is to achieve
quantitative accuracy.
In some cases reasonable trade-off between
atomistic description and analytical interpretations has emerged
from simple one dimensional two band TB models.~\cite{boykin042,boykin08}

Usually, the TB simulations are performed
within a supercell approach, which
involves periodic boundary conditions. So a
Si/barrier interface is actually modeled by a finite-width Si
slab surrounded laterally by two barriers, implying a quantum
well arrangement if the barriers are wide enough (or
superlattice for thin barriers). This introduces a width
parameter to the Si slab which was shown to play an important
role in the intervalley coupling,~\cite{grosso96,boykin041}
except for high external electric fields and wide enough Si
slabs, so that the electron interacts with a single interface.
On the other hand, within EMA it is possible to pinpoint the
role of a single interface, as both Si and the barrier material
correspond geometrically to semi-infinite slabs.

Despite these differences,
some TB studies give empirical evidence about the EMA analytical insights
obtained here:
(i) the linear  behavior of the valley splitting
with electric field, as discussed in Sec.~\ref{sec:EMA};
(ii) the relationship between the penetration of the wave-function in the
barrier material and the barrier splitting explained in Sec.~\ref{sec:numerical} and
quantified by Eqs~(\ref{eq:integraldelta}) and~(\ref{eq:integrale}) was also empirically obtained by Srinivasan
\textit{et al.}~\cite{srinivasan08} for Si electrostatic quantum dots embedded in SiGe buffers;
(iii) finally the connection between the conduction band offset and
the valley splitting demonstrated here in Fig.~\ref{fig:barrier} similar to
the TB study reported in Ref.~\onlinecite{boykin042}.

These examples illustrate how EMA studies may contribute to clarify the physics
behind phenomena numerically obtained by more detailed atomistic approaches.

Other atomistic studies include ingredients that were not implemented in our formalism, such as magnetic field and interface disorder, so that it is not possible to quantitatively compare the results demonstrated within our general EMA study to those obtained within TB for particular systems/geometries.

\subsection{Low Field Dependence of the Valley Splitting}
\label{sec:field}

 A non-linear dependence of $V_{VO}$ with $F$ has been presented in TB studies by Grosso \textit{et al.}~\cite{grosso96} and Boykin \textit{et al.}~\cite{boykin041} at low fields. We can not assess this behavior because of  instabilities in our numerical procedure, mainly due to the size of the simulation cell required by the wavefunction spread at very low electric fields. It is possible that this nonlinear behavior is connected to the quantum well behavior at low $F$, since the models in
Refs.~\onlinecite{grosso96} and~\onlinecite{boykin041}
refer to a quantum well, involving lengths associated with
the Si well and the barriers widths, on which (particularly the
Si well width) the TB results are quite sensitive.
In all cases a linear behavior is obtained at larger $F$
values. The linear behavior is predicted in Ref.\onlinecite{sham79} from Eq.~(\ref{eq:SeN}), obtained there, with $\langle\frac{\partial u}{\partial z}\rangle=F$. It also emerges from
a model for the  variational envelope in a triangular-type potential with infinite barrier potential, simpler but still similar to Eq.~(\ref{eq:trial-parts}), proposed by  Fang-Howard~\cite{fang66} and generalized by Friesen \textit{et al.}~\cite{friesen07} to allow some penetration probability into the barrier region.
In this case an analytic solution is obtained, and it can be shown that $|\Psi(z_I)|^2$ is proportional to the electric field, leading to linear increase of the intervalley coupling with $F$.
However, in most cases of interest the applied electric field should lead to a large enough $V_{VO}$, thus well within the linear regime ($F \gtrsim 10^{-3}$ V/nm here).

\section{summary and conclusions}
\label{sec:conclusions}

We presented an EMA-based study of the valley splitting in Si
induced by a Si/barrier interface. Our approach combines EMA with calculated Bloch functions, for which we give values of the relevant plane-wave expansion coefficients.
The range of splittings $2|V_{VO}|$ we obtained are in fair
agreement with measurements in Si/SiO$_2$ and Si/SiGe
interfaces.~\cite{takashina06,goswami07} Our results for $|V_vo|$   are comparable with experimentally measured values, indicating that we have probably included the most relevant physical  ingredients in our model, as discussed in Ref.~\onlinecite{saraiva09} and expanded here. In
particular, we show here that the puzzling values of the valley
splittings obtained in Ref.~\onlinecite{takashina06} could
result from particularly sharp interfaces in the very high
electric field regime. We also confirm the linear dependence of the valley splitting on the electric field in this regime. Nonetheless, many effects contributing to the valley splitting are not
included, such as strain, interface
misorientation,~\cite{friesen07} atomic scale
disorder,~\cite{kharche07} lateral
confinement,~\cite{srinivasan08} or many-body
corrections~\cite{sham79,friesen07} and the recently proposed
contribution from interface states.~\cite{saraiva10}

It is important to reiterate here the double-focus of this paper : (i) to examine the effect of an applied electric field on the valley-orbit coupling at the interface, and (ii) to provide a clear interpretation of the physics of valley splitting.  Understanding these points would hopefully reveal fundamental elements that affect the electron valley coupling and guide the identification of  a suitable environment for electron spin qubits.  By using the effective mass approximation we implicitly assume that the atomic configuration at the interface is not changed by the applied field (which should be a good approximation at low fields), and that interface roughness is small at the length scale over which the electron wave function changes significantly.~\cite{culcer10b} A comprehensive study of valley splitting, including the effects of interface roughness, thickness, applied electric field, and other elements mentioned in the previous paragraph, would necessarily contain an atomistic component. Such a study would be required to clearly establish the limit of applicability of EMA, which is essentially a mean field approach for the bound electron wave function, interface disorder and composition profiles.

Our EMA approach provides a clear interpretation of the physics
of valley splitting, revealing fundamental elements affecting
the coupling and eventually providing a suitable environment
for electronic spin qubit operation. While the investigation of
other effects mentioned above is desirable, a simpler model
that goes beyond the phenomenological approach, as provided
here, is useful in guiding nanofabrication and device operation
efforts. In particular, a profound understanding of the valley coupling induced by interfaces may pave the road to the quantum manipulation and processing of the valley degree of freedom.~\cite{culcer11} 

We also provide a critical analysis of the valley physics theory available in the literature, and discuss some of the ingredients that are imperative in a comprehensive theory. These include the correct interpretation of the valley coupling as a complex number and the inclusion of Umklapp processes. 

In conclusion, we have calculated electron valley splitting in
Si at a Si/barrier interface. We show that a sizeable
single-particle valley-orbit coupling can be obtained by applying a high
enough external field and choosing an optimal barrier material
providing a suitable potential barrier height and high quality
abrupt interface.

\begin{acknowledgments}
We thank Mark Friesen, Kei Takashina and
Yukinori Ono for helpful and fruitful discussions. This work was partially supported by the Brazilian agencies
CNPq, FUJB, FAPERJ, and performed  as part of the INCT on Quantum
Information/MCT. MJC
acknowledges FIS2009-08744 and the Ram\'{o}n y Cajal program
(MICINN, Spain). XH and SDS thank financial support by NSA and
LPS through US ARO. 
\end{acknowledgments}
\bibliography{long_vs}

\end{document}